\documentclass[12pt]{article}
\usepackage[dvips]{graphicx}
\usepackage{pdproc}

  \makeatletter 
  \def\@cite#1{[#1]} 
  \makeatother    
\begin{document}

\renewcommand{\thefootnote}{\alph{footnote}}

\title{
Resonant Higgs-Sector CP Violation at the LHC
}

\author{ 
JAE SIK LEE
}

\address{ 
Department of Physics and Astronomy, University of Manchester\\
Manchester M13 9PL, United Kingdom \\
}

\abstract{
We present  the general formalism for  studying CP-violating phenomena
in  the  production,   mixing  and  decay  of  a   coupled  system  of
CP-violating   neutral   Higgs   bosons  at   high-energy   colliders.
Considering  the Minimal  Supersymmetric Standard  Model  (MSSM) Higgs
sector in which  CP violation is radiatively induced  by phases of the
soft   supersymmetry-breaking    parameters,
we  apply our formalism to neutral Higgs
production via ${\bar b}b$, $gg$  and $W^+ W^-$ collisions and their
subsequent decays into $\tau^+\tau^-$ pairs at the LHC.
We  discuss   CP  asymmetries  in  the   longitudinal  
polarizations  of  $\tau^+  \tau^-$   pairs 
based on a scenario in which
all neutral   Higgs  bosons   are  nearly  degenerate   and the CP-violating
signatures are resonantly enhanced.
}

\normalsize\baselineskip=15pt

\section{Introduction}
In the presence of non-vanishing CP phases in the soft SUSY-breaking terms,
radiative corrections naturally induce a mixing between CP-odd and CP-even
Higgs states \cite{Earlier}.
The main contribution comes from stop and sbottom loops due to the large
Yukawa couplings, the size of the CP-violating mixing being proportional to
\cite{Later}
\begin{equation}
\Delta_{\tilde{f}}\equiv\frac{\Im{\rm m}(A_f\,\mu)}
{(m_{\tilde{f}_2}^2-m_{\tilde{f}_1}^2)}
\end{equation}
with $f=t,b$. Here the $A_f$'s are complex trilinear soft-breaking parameters,
$\mu$ is the complex supersymmetric Higgs(ino) mass parameter, and
the $m_{\tilde{f}_i}$, $i=1,2$, are the squark mass eigenvalues.
Even though the mixing is loop-suppressed, its effects 
can be sizable even for small values of $\Im{\rm m}(A_f\,\mu)$,
especially when two or more neutral Higgs bosons are nearly degenerate
\cite{ELP}.

In the presence of this non-trivial mixing with non-vanishing CP phases, the
neutral Higgs bosons do not have to carry any definite CP parity and the
mixing among them is described by the orthogonal 3$\times$3
matrix instead of a 2$\times$2 one
and, accordingly, the couplings of the Higgs bosons to the SM and 
SUSY particles are greatly modified. 
The most prominent example is the Higgs-boson coupling to a
pair of vector bosons, which is responsible for the production of Higgs
bosons at $e^+e^-$ colliders. In fact, the lightest Higgs boson $H_1$ lighter
than 50 GeV could have been undetected at LEP2 for intermediate values of
$\tan\beta$ \cite{OPAL}.

Recently a computational tool called CPsuperH \cite{CPsuperH}
has been developed for Higgs
phenomenology in the MSSM with explicit CP violation. This code calculates the
Higgs boson pole masses \cite{PoleMass},
Higgs-boson mixing matrix, all the couplings of Higgs
bosons to the SM and SUSY particles, and the decay widths and branching
fractions of the neutral and charged Higgs bosons \cite{Decays}. 

\section{Higgs-boson propagator matrix}

In a situation where two or more neutral Higgs bosons are simultaneously 
involved in a process due to their  mass
differences comparable  to their  decay widths, the full $3\times 3$ Higgs
boson propagator matrix $D(\hat{s})$ should be considered
\footnote{Here we neglect the small off-diagonal absorptive parts between 
Goldstone and Higgs bosons. }:

\newcommand{\imag}{\Im {\rm m}}
\newcommand{\real}{\Re {\rm e}}
\begin{equation}
  \label{eq:hprop}
D (\hat{s}) = \hat{s}\,
\left(\begin{array}{ccc}
       \hat{s}-M_{H_1}^2+i\imag\widehat{\Pi}_{11}(\hat{s}) &
i\imag\widehat{\Pi}_{12}(\hat{s})&
       i\imag\widehat{\Pi}_{13}(\hat{s}) \\
       i\imag\widehat{\Pi}_{21}(\hat{s}) &
\hat{s}-M_{H_2}^2+i\imag\widehat{\Pi}_{22}(\hat{s})&
       i\imag\widehat{\Pi}_{23}(\hat{s}) \\
       i\imag\widehat{\Pi}_{31}(\hat{s}) & i\imag\widehat{\Pi}_{32}(\hat{s}) &
       \hat{s}-M_{H_3}^2+
       i\imag\widehat{\Pi}_{33}(\hat{s})
      \end{array}\right)^{-1} \,,
\end{equation}
where $M_{H_i}$'s are Higgs-boson pole masses and for the absorptive parts  of
the Higgs-boson  propagator matrix,  $i\imag\widehat{\Pi}_{ij}(\hat{s})$, we
have included contributions from loops of  
fermions (third generation leptons and quarks, charginos, and neutralinos), 
vector bosons, associated pairs
of Higgs and vector bosons, Higgs-boson pairs, and sfermions.

\begin{figure}[htb]
\begin{center}
\includegraphics*[width=10cm]{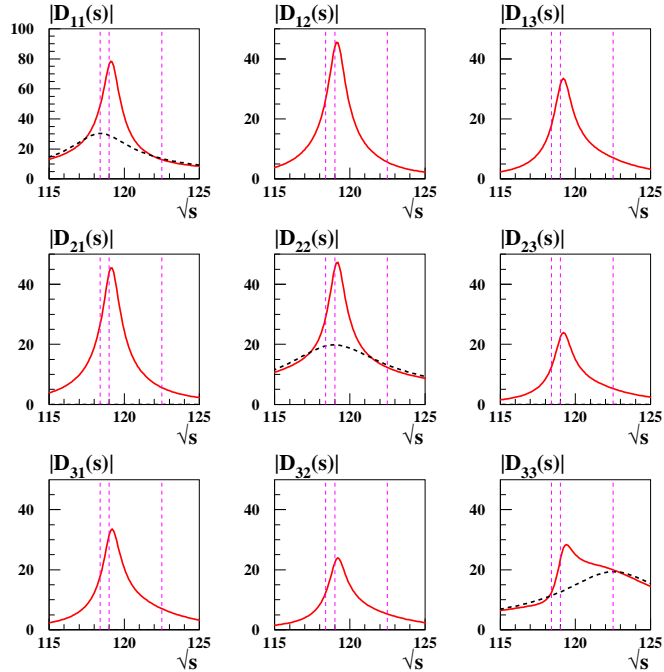}
\caption{%
The absolute value of each component of the Higgs-boson propagator matrix
$|D_{ij}(s)|$ with (red solid lines) and without (black dashed lines)
including off-diagonal absorptive parts
in the three-way mixing scenario
with $\Phi_{A}=-\Phi_3=90^\circ$. 
Three Higgs-boson pole masses are presented with vertical lines.
}
\label{fig:dh3mix}
\end{center}
\end{figure}

In Fig.~\ref{fig:dh3mix}, we show the absolute value of each component of the 
Higgs-boson propagator matrix, $|D_{ij}(s)|$, in the so-called three-way mixing
scenario, which is characterized by a large value of $\tan\beta=50$ and 
small $M_{H^{\pm}}^{\rm pole}=155$ GeV
with $\Phi_{A}=-\Phi_3=90^\circ$ \cite{ELP};
$\Phi_{A}\equiv{\rm arg}(A_{t}\mu)={\rm arg}(A_{b}\mu)$ and
$\Phi_{3}\equiv{\rm arg}(M_3\mu)$ with $M_3$ the gluino mass parameter. 
We observe the inclusion of the off-diagonal
absorptive parts significantly modifies the Higgs-boson propagators: the peak
positions could be different from pole-mass positions and the off-diagonal
transition propagators can be sizable and should not be neglected.

\section{At the LHC}
 
We have considered three main neutral
Higgs-boson production mechanisms at the LHC
and the significant mixing among Higgs
bosons before decaying into $\tau$ leptons:
\begin{itemize}
\item $b$-quark fusion : 
$pp (b\bar{b}) \rightarrow H_{i\rightarrow j} X \rightarrow \tau^+\tau^- X$ 
\item gluon fusion :
$pp (gg) \rightarrow H_{i\rightarrow j} X \rightarrow \tau^+\tau^- X$ 
\item $W$-boson fusion:
$pp (W^+W^-) \rightarrow H_{i\rightarrow j} X \rightarrow \tau^+\tau^- X$ 
\end{itemize}
%
\begin{figure}[htb]
\begin{center}
\includegraphics*[width=10cm]{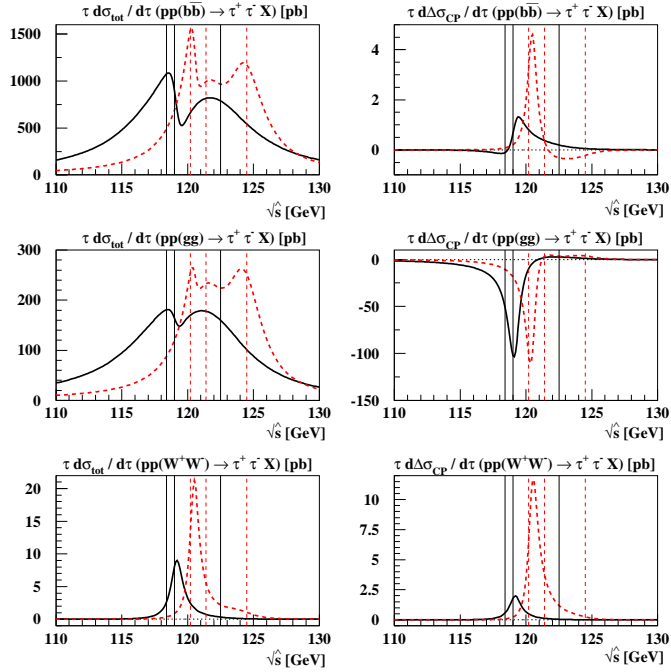}
\caption{%
The differential total and CP-violating cross sections,
$\tau \frac{d\sigma_{\rm tot}}{d\tau}$ and
$\tau \frac{d\Delta\sigma_{\rm CP}}{d\tau}$,
as functions of the
$tau$-lepton invariant mass $\sqrt{\hat{s}}=\sqrt{(p_{\tau^+}+p_{\tau^-})^2}$
in the three-way mixing scenario with
$(\Phi_{A},\Phi_3)=(90^\circ,-90^\circ)$ (black solid lines)  and
$(\Phi_{A},\Phi_3)=(90^\circ,-10^\circ)$ (red dashed lines).  The vertical
lines show the positions of three Higgs-boson pole masses in each case.
The kinematic parameter 
$\tau=\hat{s}/s$ with $s$ the hadron-collider energy squared.
}
\label{fig:s12hp}
\end{center}
\end{figure}

In Fig.~\ref{fig:s12hp}, we show
the differential total and CP-violating cross sections defined by :
\begin{eqnarray}
\sigma_{\rm tot}&\equiv &\sigma (pp \rightarrow \tau^+_R\tau^-_R X)
                +\sigma (pp \rightarrow \tau^+_L\tau^-_L X)\,, \nonumber \\
\Delta\sigma_{\rm CP} &\equiv &\sigma (pp \rightarrow \tau^+_R\tau^-_R X)
                -\sigma (pp \rightarrow \tau^+_L\tau^-_L X)\,.
\end{eqnarray}
It is indispensable to measure the longitudinal polarizations of tau leptons
to construct the CP-violating cross section $\Delta\sigma_{\rm CP}$ which is
non-vanishing only in the presences of non-trivial CP phases and the
absorptive parts in the Higgs-boson propagators or their vertices.

We observe that
the $b$-quark fusion is the dominant production mechanism \cite{BLS}
and the total cross section is at least five times larger than the
gluon fusion cross section. This large cross-section reduces the 
would-be large CP asymmetry in the gluon-fusion channel below 1\% level 
after combining $b$-quark and gluon fusion channels. 
The reason is that in the scenario under consideration $b$-quark rescattering
effects dominate the resonant transition amplitude, leading to a strong
suppression of CP violation \cite{ELP}.
We also note that the peak positions don't exactly coincide with the
Higgs-boson pole-mass positions and the resonance shape strongly depends 
on the production channels.

Though the WW-fusion cross section is the smallest
one, this process can be discriminated experimentally from
$b$-quark and gluon fusion ones and can be
very promising for detecting Higgs
sector CP violation at the LHC. We find that the CP asymmetry given by the ratio
$\sigma_{\rm tot}/\Delta\sigma_{\rm CP}$ is large over wide range of CP
phases. Especially, it can be easily larger than 30 \% even 
for small CP phases and
reach to 80 \% when $\Phi_3=-70^\circ$ as shown in Figs.~8--11 of
Ref.~\cite{ELP}.

\section{Acknowledgements}
J.S.L would like to thank John Ellis and Apostolos Pilaftsis for
collaborations. Work supported in part by PPARC.

\bibliographystyle{plain}

\end{document}